\documentclass[preprints,article,accept,moreauthors,pdftex]{Definitions/mdpi}

\usepackage{hyperref}
\makeatletter
\def\UrlAlphabet{%
\do\a\do\b\do\c\do\d\do\e\do\f\do\g\do\h\do\i\do\j%
\do\k\do\l\do\m\do\n\do\o\do\p\do\q\do\r\do\s\do\t%
\do\u\do\v\do\w\do\x\do\y\do\z\do\A\do\B\do\C\do\D%
\do\E\do\F\do\G\do\H\do\I\do\J\do\K\do\L\do\M\do\N%
\do\O\do\P\do\Q\do\R\do\S\do\T\do\U\do\V\do\W\do\X%
\do\Y\do\Z}
\def\UrlDigits{\do\1\do\2\do\3\do\4\do\5\do\6\do\7\do\8\do\9\do\0}
\g@addto@macro{\UrlBreaks}{\UrlOrds}
\g@addto@macro{\UrlBreaks}{\UrlAlphabet}
\g@addto@macro{\UrlBreaks}{\UrlDigits}
\makeatother

\firstpage{1} 
\pubvolume{xx}
\issuenum{1}
\articlenumber{5}
\pubyear{2020}
\copyrightyear{2020}
\history{Received: 14 September 2020; Accepted: 3 October 2020; Published: date}

\Title{Utilising Deep Learning Techniques for Effective Zero-Day Attack Detection} 

\Author{Hanan~Hindy~$^{1,}$*\orcidA{},  
Robert~Atkinson~$^{2}$\orcidB{},
Christos~Tachtatzis~$^{2}$\orcidC{},
Jean-No\"{e}l~Colin~$^{3}$\orcidD{},
Ethan~Bayne~$^{1}$\orcidE{}, and
Xavier~Bellekens~$^{2}$\orcidF{},
}

\AuthorNames{Hanan Hindy, Robert Atkinson, Christos Tachtatzis Jean-No\"{e}l Colin, Ethan Bayne and Xavier Bellekens}

\address{%
$^{1}$ \quad Division of Cybersecurity, Abertay University, Dundee DD1 1HG, UK; {1704847@abertay.ac.uk (H.H.); e.bayne@abertay.ac.uk (E.B)}\\
$^{2}$ \quad Electronic and Electrical Engineering Department, 
 University of Strathclyde, Glasgow, G1 1XQ, UK; 
 {robert.atkinson@strath.ac.uk (R.A.); christos.tachtatzis@strath.ac.uk (C.T.); xavier.bellekens@strath.ac.uk (X.B.)}\\
$^{3}$ \quad InfoSec Research Team, University of Namur, 5000 Namur, 
 Belgium; jean-noel.colin@unamur.be}

\corres{Correspondence: hananhindy@ieee.org}

\abstract{Machine Learning~(ML) and Deep Learning~(DL) have been used for building Intrusion Detection Systems~(IDS). The~increase in both the number and sheer variety of new cyber-attacks poses a tremendous challenge for IDS solutions that rely on a database of historical attack signatures. 
Therefore, the~industrial pull for robust IDSs that are capable of flagging zero-day attacks is growing. Current~outlier-based zero-day detection research suffers from high false-negative rates, thus limiting their practical use and performance. This paper proposes an autoencoder implementation for detecting zero-day attacks. The~aim is to build an IDS model with high recall while keeping the miss rate (false-negatives) to an acceptable minimum. Two well-known IDS datasets are used for evaluation---CICIDS2017 and NSL-KDD. In order to demonstrate the efficacy of our model, we compare its results against a One-Class Support Vector Machine~(SVM). The~manuscript highlights the performance of a One-Class SVM when zero-day attacks are distinctive from normal behaviour. The~proposed model benefits greatly from autoencoders encoding-decoding capabilities. The~results show that autoencoders are well-suited at detecting complex zero-day attacks. The~results demonstrate a zero-day detection accuracy of [89--99\%] for the NSL-KDD dataset and [75--98\%] for the CICIDS2017 dataset. Finally, the~paper outlines the observed trade-off between recall and fallout. 
}

\keyword{\hspace{-8mm} Autoencoder;~Artificial Neural Network;~One-Class Support Vector Machine; Intrusion~Detection; Zero-day Attacks; CICIDS2017; NSL-KDD}

\usepackage{booktabs} 
\usepackage{multirow}
\usepackage{soul} 
\usepackage{microtype}
\graphicspath{{Definitions/}}

\usepackage{xcolor, colortbl}
\usepackage[labelformat=simple]{subcaption}

\DeclareCaptionLabelFormat{subcaptionlabel}{\normalfont(\textbf{#2}\normalfont)}
\usepackage{amsmath}
\usepackage[utf8]{inputenc}
\inputencoding{utf8}

\usepackage{comment}
\usepackage{hyperref}

\usepackage{bm}
\DeclareMathOperator*{\argminB}{argmin}

\usepackage{algorithm}
\usepackage{algpseudocode}

\usepackage{array}
\newcolumntype{C}[1]{>{\centering\arraybackslash}m{#1}}
\newdimen\NetTableWidth
 \NetTableWidth=\dimexpr
    \linewidth - 8\tabcolsep - 5\arrayrulewidth

\makeatletter
\newcommand{\StatexIndent}[1][3]{%
  \setlength\@tempdima{\algorithmicindent}%
  \Statex\hskip\dimexpr#1\@tempdima\relax}
\algdef{S}[WHILE]{WhileNoDo}[1]{\algorithmicwhile\ #1}%
\makeatother

\usepackage{soul}
\soulregister\ref7
\soulregister\cite7

\usepackage{mathtools}
\DeclarePairedDelimiter{\norm}{\lVert}{\rVert}

\makeatletter
\newcount\SOUL@minus 
\makeatother

\begin{document}


\section{Introduction}
\label{sec:introduction}

Central to tackling the exponential rise in cyber-attacks~\cite{10.1145/3372823, 8551386}, is Intrusion Detection Systems~(IDS) systems that are capable of detecting zero-day cyber-attacks. 
Machine Learning~(ML) techniques have been extensively utilised for designing and building robust IDS~\cite{Khraisat2019, HananHindy2018}. However, while current IDS can achieve high detection accuracy for known attacks, they often fail to detect new, zero-day attacks. This is due to the limitations of current IDS, which~rely on pre-defined patterns and signatures. Moreover, current IDS suffer from high false-positive rates, thus limiting the performance and their practical use in real-life deployments. \mbox{As a result,} large numbers of zero-day attacks remain undetected, which escalate their consequences (denial of service, stolen customer details, etc.). 

According to Chapman~\cite{CHAPMAN20161}, a zero-day attack is defined as "a traffic pattern of interest that, in general, has no matching patterns in malware or attack detection elements in the network"~\cite{CHAPMAN20161}.
The~implications of zero-day attacks in real-world are discussed by Bilge and Dumitras~\cite{10.1145/2382196.2382284}. Their research focuses on studying their impact and prevalence. The~authors highlight that zero-day attacks are significantly more prevalent than suspected, demonstrating that, out of their 18 analysed attacks, 11 (61\%) were previously unknown~\cite{10.1145/2382196.2382284}. Their findings showed that a zero-day attack can exist for a substantial period of time (average of 10 months~\cite{10.1145/2382196.2382284}) before they are detected and can compromise systems during that period. Additionally, Nguyen and Reddi~\cite{nguyen2019deep} refer to a statistical study that shows that 62\% of the attacks are identified after compromising systems.
Moreover, the~number of zero-day attacks in 2019 exceeds the previous three years~\cite{ZeroDayE16:online}. All of these considerations highlight the clear and urgent need for more effective attack detection models.

One of the main research directions to detect zero-day attacks relies on detecting outliers (i.e.,~instances/occurrences that vary from benign traffic). However, the~main drawback of the available outlier-based detection techniques is their relatively low accuracy rates as a result of both high false-positive rates and high false-negative rates. As discussed, the~high false-negative rates leave the system vulnerable to attack, while the high false-positive rates needlessly consume the time of cyber security operation centres; indeed, only 28\% of investigated intrusions are real~\cite{cisco2017}. Ficke~{et al.}~\cite{9020860} emphasise the limitations that false-negative could bring to IDS development, for example, it reduces IDS effectiveness. 

Sharma~{et al.}~\cite{sharma2018framework} proposed a framework to detect zero-day attacks in Internet of Things~(IoT) networks. They rely on a distributed diagnosis system for detection. Sun~{et al.}~\cite{RefWorks:doc:5c812934e4b075899776c5dd} proposed a Bayesian probabilistic model to detect zero-day attack paths. The~authors visualised attacks in a graph-like structure and introduced a prototype to identify  attacks. 
Zhou and Pezaros~\cite{zhou2019evaluation} evaluated six different supervised ML techniques; using the CIC-AWS-2018 dataset. The~authors use decision tree, random forest, k-nearest neighbour, multilayer perceptron, quadratic discriminant analysis, and gaussian na\"ive bayes classifiers. The~authors do not fully detail how these supervised ML techniques are trained on benign traffic solely to be utilised for unknown attacks detection or how zero-day (previously unseen) attacks are simulated and detected. Moreover, transfer learning is used to detect zero-day attacks. Zhao~{et al.}~\cite{zhao2019transfer} used transfer learning to map the connection between known and zero-day  attacks~\cite{zhao2019transfer}. Sameera and Shashi~\cite{SAMEERA2020} used deep transudative transfer learning to detect zero-day attacks. 

Furthermore, ML is used to address Zero-day malware detection. For~example, Abri~{et al.} evaluated the effectiveness of using different ML techniques (Support Vector Machine~(SVM), Na\"ive Bayes, Multi-Layer Perceptron, Decision trees, k-Nearest Neighbour, and Random Forests) to detect zero-day malware~\cite{RefWorks:doc:5dde4cdee4b0518003692b09}, while Kim~{et al.}~\cite{RefWorks:doc:5ddeb66fe4b0227c51200fde} proposed the  use of Deep-Convolutional Generative Adversarial Network~(DCGAN).

In this paper, we propose utilising the capabilities of Deep Learning~(DL) to serve as outlier detection for zero-day attacks with high recall. The~main goal is to build a lightweight intrusion detection model that can detect new (unknown) intrusions and zero-day attacks, with a high recall (true-positive rate) and low fallout (false-positive rate).
Accordingly, having a high detection capability of zero-day attacks will help to reduce the complications and issues that are associated with new attacks.

The contributions of this work are threefold;
\begin{itemize}[leftmargin=*,labelsep=4.9mm]%
    \item Proposing and implementing an original and effective autoencoders model for zero-day detection~IDS.
    \item Building an outlier detection One-Class SVM model.
    \item Comparing the performance of the One-Class SVM model as a baseline outlier-based detector to the proposed Autoencoder model.
\end{itemize}

The rest of the paper is organised as follows; the background is presented in Section~\ref{sec:background}, Section~\ref{sec:related-work} discusses the related work showing the results and approaches of recent IDS research. Section~\ref{sec:datasets} describes the datasets that are used and how zero-day attacks are simulated. In Section~\ref{sec:proposed-models}, the~proposed models are explained. Section~\ref{sec:results} presents the experimental results and findings. Finally, the~paper is concluded in Section~\ref{sec:conclusion}.

\section{Background}
\label{sec:background}
In this section, the~models utilised in this investigation are discussed. Section~\ref{sec:background:autoencoders} describes the deep-learning based autoencoder model and Section~\ref{sec:background:SVM} describes an unsupervised variant of a support vector machine model. 

\subsection{Autoencoders} 
\label{sec:background:autoencoders}
The model that is proposed in this manuscript principally benefits from the autoencoder characteristics and attributes. The~objective is that the autoencoder acts as a light-weight outlier detector, which could then be used for zero-day attacks detection, as further discussed in Section~\ref{sec:autoencoder}.

\begin{figure}[H]
    \centering
    \includegraphics[width=0.5\linewidth]{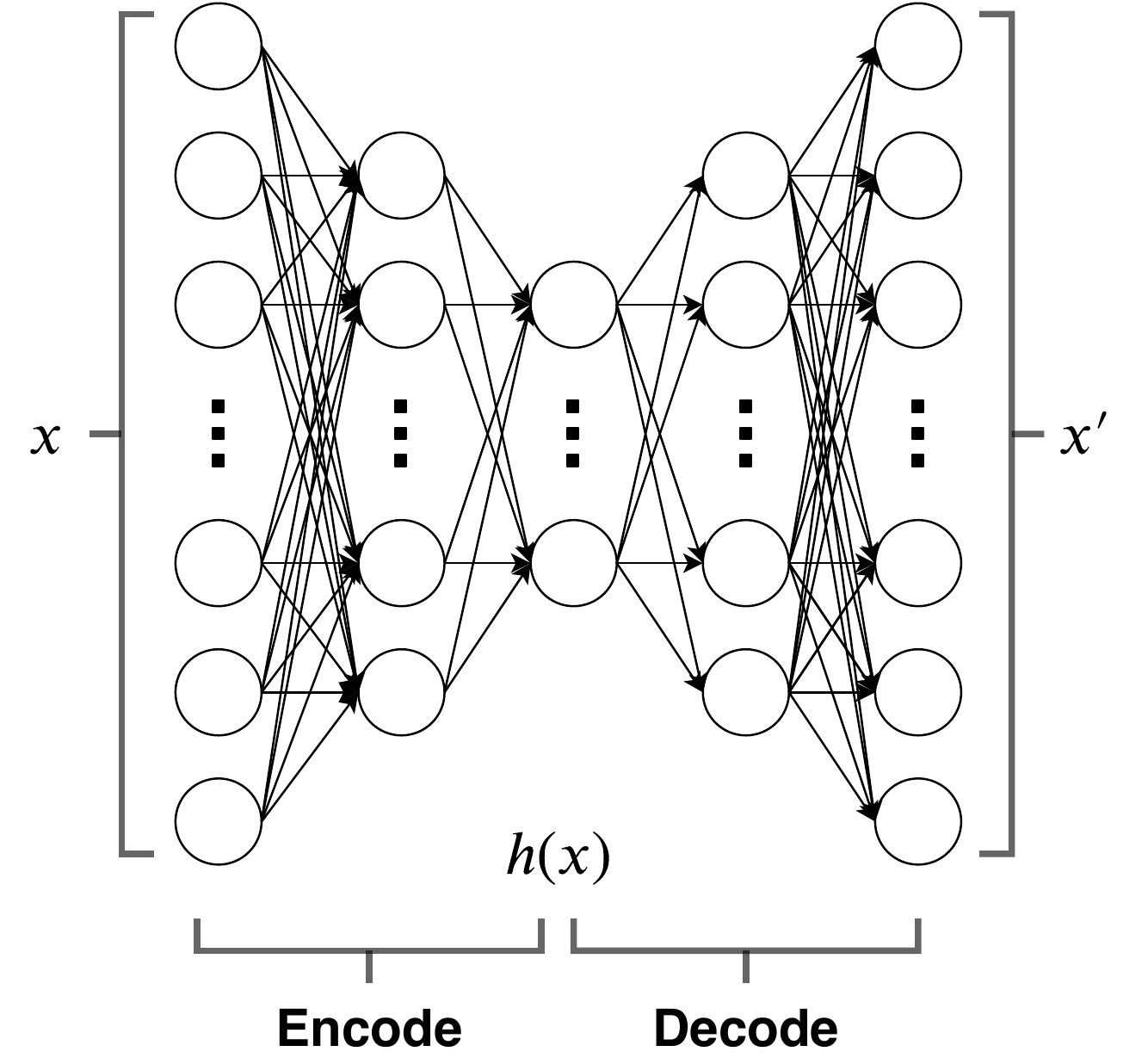}
    \caption{Autoencoder Architecture}
    \label{fig:autoencoder}
\end{figure}

Rumelhart~{et al.}~\cite{RefWorks:doc:5dd2bfdfe4b0a03ea81aa826} first introduced autoencoders in order to overcome the back propagation in unsupervised context using the input as the target. Autoencoders are categorised as self-supervised, since the input and the output are particularly the same~\cite{Comprehe85}.
As defined by Goodfellow~{et al.}~\cite{RefWorks:doc:5dd2b90be4b0a03ea81aa557}, an~Autoencoder is ``a neural network that is trained to attempt to copy its input to its output''~
\cite{RefWorks:doc:5dd2b90be4b0a03ea81aa557}. Figure~\ref{fig:autoencoder} illustrates the basic architecture of an autoencoder. The~architecture of an autoencoder and the number of hidden layers differ based on the domain and the usage scenario. 

Formally, given an input $\mathcal{X}$, an autoencoder is trained in order to minimise the reconstruction error, as demonstrated in Equation~\eqref{eq:minimisation}~\cite{Comprehe85}.
\begin{equation}
    \centering  
    \label{eq:minimisation}
    \begin{gathered}
    \bm{\phi : \mathcal{X} \rightarrow \mathcal{F}} \\
    \bm{\psi : \mathcal{F} \rightarrow \mathcal{X}} \\
    \bm{\phi, \psi = \argminB_{\phi, \psi} ||\mathcal{X} - (\phi \circ \psi)\mathcal{X}||^2}
    \end{gathered} 
    \qquad 
\end{equation}

\noindent
such that $\phi$ and $\psi$ represent the encoding and decoding functions, respectively.

Commonly, the~reconstruction error of an input $x$ is represented as the difference between $x$ and $x'$, such that:
\vspace{6pt}

$x' = g(f(x))$ 

\vspace{6pt}

\noindent where $f(x)$ is the encoding function, constructing the encoded vector of $x$

\noindent $g(x)$ is the decoding function, restoring $x$ to its initial value\\ 

The reconstruction error is defined by a function that represents the difference between the input $x$ and the reconstructed input $x'$. Mean square error and mean absolute error are common functions that are used in order to calculate the reconstruction error, as shown in Equations~\eqref{eq:mse} and \eqref{eq:mae}, respectively.
\begin{equation}
    \centering
    \label{eq:mse}
    MSE = \sum_{i=1}^{N}{(x'-x)^2}
\end{equation}
\begin{equation}
    \centering
    \label{eq:mae}
    MAE = \sum_{i=1}^{N}{|x'-x|}
\end{equation}

Supposing that the encoding function $f(x)$ is single layer network with a linear function, the~Autoencoder is viewed as equivalent to Principal Components Analysis~(PCA)~\cite{barber2014implicit}.

Autoencoders were originally used for dimensionality reduction and feature learning~\cite{RefWorks:doc:5dd2b992e4b0ff5398154ceb, RefWorks:doc:5dd2b944e4b0a03ea81aa581}. However, many other applications have been recently proposed. These applications include: word semantics~\cite{RefWorks:doc:5dd2ba6ee4b039d5609d627e}, image compression~\cite{RefWorks:doc:5dd2bd88e4b0caca97c355e0}, image anomaly detection~\cite{RefWorks:doc:5dd2bc1ae4b0c0c541dadb08}, denoising~\cite{RefWorks:doc:5dd2b844e4b064aae78e03e6}, and others.

\subsection{One-Class SVM} \label{sec:background:SVM}
The SVM is one of the most well-established supervised ML techniques. Given the training samples, an SVM is trained to construct a hyperplane in a high-dimensional space that best separates the classes~\cite{cortes1995support}. This hyperplane is a line in two dimensions case, a plane in the case of three dimensions (3D) or higher dimensions (n-dimensional).
When data are not linearly separable, a kernel is used to map the input features/data to a non-linear higher dimensional space, in which a hyperplane would best separate the classes. The SVM kernels include; linear, polynomial, Gaussian, and Radial Basis Function~(RBF).

Formally, given two classes, the~minimisation problem of SVM is represented, as shown in Equation~\eqref{eq:svm-min}~\cite{ng2000cs229}.
\begin{equation}
\label{eq:svm-min}
    \begin{gathered}
    \min_{\textrm{w}\in \mathbb{R}^d} \norm{\textrm{w}}^2 + C \sum_{i}^{N}{\textrm{max}(0, 1 -  y_if(x_i))} \\
    f(x_i) = (\textrm{w}^T x_i + b)
     \end{gathered}
     \qquad
\end{equation}
where $C$ is a regularisation parameter to represent the trade-off between ensuring that $x_i$ is on the expected side of the plane and increasing the margin. 
Based on Equation~\eqref{eq:mae}, the~data points fall in one of three places based on $ y_if(x_i)$. If $y_if(x_i)$ is greater than 1, then the point is outside the margin and it does not contribute to the loss. If $y_if(x_i)$ equals 1, then the point is on the margin. Finally, if $y_if(x_i)$ is less than 1, then the point contributes to the loss, as it is on the wrong side~\cite{Hilary2015}.

In contrast to its supervised counterpart, the~One-Class SVM is an unsupervised variant. It is defined as a model that is capable of detecting ``Novelty''~\cite{scholkopf2000support}. The~goal of One-Class SVM is to fit a hyperplane that acts as a boundary which best includes all the training data and excludes any other data point. The~result of training a One-Class SVM is seen as a spherically shaped boundary~\cite{tax2004support}. Because One-Class SVM is considered to be one of the most established outlier-based ML techniques, it provides an ideal comparison for assessing the performance of a deep neural network based autoencoder. 

Formally, given a class with instances $\{x_1, ...., x_N\}$, and a mapping function $\varphi()$ that maps the features to a space $H$, the~goal of One-Class SVM is to fit a hyperplane $\Pi$ in $H$ that has the largest distance to the origin, and all $\varphi(x_i)$ lie at the opposite side of hyper-plane to the origin~\cite{WANG2018198}.

\section{Related Work}
\label{sec:related-work}

IDS is defined as "a system or software that monitors a network or systems for malicious activity". Generally, IDSs can either be Network Intrusion Detection System~(NIDS) or Host Intrusion Detection System~(HIDS). NIDS monitors the network and communication while HIDS monitors the internal operation and log files~\cite{hodo2017shallow}. Based on their detection techniques, IDSs are classified into Signature-based IDS, which~relies on known signatures of prior and known attacks, and Anomaly-based IDS, which~relies on patterns~\cite{RefWorks:doc:5cdd85f5e4b02cf23c27b848}. When compared to signature-based IDS, anomaly-based IDS perform better with complex attacks and unknown attacks.

In the past decade, researchers developed multiple techniques in order to enhance the robustness of anomaly-based IDS. Subsequent to a long period of using statistical methods to detect cyber anomalies and attacks, the~need for ML emerged. 
Because of the sophistication of cyber-attacks, statistical methods were rendered inadequate to handle their complexity. 
Therefore, with the advancement of ML and DL in other domains (i.e.,~image and video processing, natural language processing, etc.), the researchers adopted these techniques for cyber use. Nguyen and Reddi~\cite{nguyen2019deep} discuss the importance and benefit ML can provide to cybersecurity by granting a `robust resistance' against attacks.

Based on the analysis of recent IDS research~\cite{9108270}, ML has dominated the IDS research in the past decade. The~analysis shows that Artificial Neural Networks (ANN), SVM, and k-means are the prevailing algorithms. Buczak and Guven~\cite{7307098} analyse the trends and complexity of different  ML and DL techniques used for IDS.
Moreover, recent research is directed towards the use of DL to analyse network traffic, due to the DL capabilities of handling complex patterns. {Due to the complexity and challenges that are associated with encrypted traffic, building robust and reliable DL-based IDSs is crucial. Aceto~et al.~\cite{ACETO2020306} describe this advancement in traffic encryption, as it `defeats traditional techniques' that relies on packet-based and port-based data.}
At the beginning of 2019, 87\% of traffic was encrypted~\cite{Encrypti14}, which~emphasises on the growth and, thus the need for corresponding IDs. Research has utilised flow-based features as the building block for training and analysing IDSs in order to handle encrypted and non-encrypted traffic. The~benefit of flow-based features, when compared to packet-based ones, relies on the fact that they can be used with both encrypted and unencrypted traffic and also they characterise high-level patterns of network communications. New DL approaches have recently been used to build robust and reliable IDS. 
One of these techniques is autoencoders.

In the cyber security domain, autoencoders are used for feature engineering and learning. Kunang~{et al.}~\cite{8605181} used autoencoders for feature extraction, features are then passed into an SVM for classification. KDD Cup'99 and NSL-KDD datasets are both used for evaluation. The~evaluation of the model, using autoencoder for feature extraction and SVM for multi-class classification, has an overall accuracy of 86.96\% and precision of 88.65\%. The~different classes accuracies show a poor performance, as follows: 97.91\%, 88.07\%, 12.78\%, 8.12\%, and 97.47\% for DoS, probe, R2L, U2R, and normal, respectively; a precision of 99.45\%, 78.12\%, 97.57\%, 50\%, and 81.59\% for DoS, probe, R2L, U2R, and normal, respectively. 

Kherlenchimeg and Nakaya~\cite{RefWorks:doc:5f3a7005e4b07d1d33a62c96} use a sparse autoencoder to extract features. The~latent representation (the bottleneck layer of the autoencoder) is fed into a Recurrent Neural Network~(RNN) for classification. The~accuracy of the IDS model while using the NSL-KDD dataset is 80\%. Similarity, Shaikh and Shashikala~\cite{shaikh2019autoencoder} focus on the detection of DoS attacks. They utilise a Stacked Autoencoder with an LSTM network for the classification. Using the NSL-KDD dataset, the~overall detection accuracy is 94.3\% and a false positive rate of 5.7\%. 

Abolhasanzadeh~\cite{7288799} used autoencoders for dimensionality reduction and the extraction of bottleneck feature. The~experiments were evaluated while using the NSL-KDD dataset. In a similar fashion, Niyaz~{et al.}~\cite{javaid2016deep} used autoencoders for unsupervised feature learning. They used the NSL-KDD dataset. 
Additionally, AL-Hawawreh~{et al.}~\cite{RefWorks:doc:5b0e81c9e4b01f2c3e37bf75} used deep autoencoders and trained them on benign traffic in order to deduce the most important feature representation to be used in their deep feed-forward ANN. 

Shone~{et al.}~\cite{RefWorks:doc:5ae339c4e4b0e00594a5c5cc} use a Stacked Non-Symmetric Deep Autoencoder to refine and learn the complex relationships between features that are then used in the classification using random forest technique. The~authors used both the KDD Cup'99 and NSL-KDD datasets. 

Farahnakian and Heikkonen~\cite{8323687} used a deep autoencoder to classify attacks. The~deep autoencoder is fed into a single supervised layer for classification. Using the KDD Cup'99 dataset, the~highest accuracies are 96.53\% and 94.71\% for binary and multi-class classification respectively.

In agreement with our manuscript, Meidan~et al.~\cite{meidan2018n} utilised the encoding-decoding capabilities of autoencoders to learn normal behaviour in IoT networks setup. Subsequently, IoT botnet and malicious behaviour are detected when the autoencoder reconstruction fails. {Bovenzi~et al.~\cite{bovenzihierarchical} emphasised the need for adaptive ML models to cope with the heterogeneity and unpredictability of IoT networks. The~authors propose a two-stage IDS model, where they leverage the autoencoder capabilities in the first stage of their IDS.}

\section{Datasets}
\label{sec:datasets}

Two mainstream IDS datasets are chosen in order to evaluate the proposed models.
The first is the CICIDS2017 dataset~\cite{RefWorks:doc:5b227bb8e4b07f83f15ddb45}, which was developed by the Canadian Institute for Cybersecurity~(CIC). The~CICIDS2017 dataset covers a wide range of recent insider and outsider attacks. It comprises a diverse coverage of protocols and attacks variations and, finally, it is provided in a raw format which enables researchers the flexibility of processing the dataset. Therefore, the~CICIDS2017 dataset is well-suited for evaluating the proposed models.

The CICIDS2017 dataset is a recording of a five-day benign, insider and outsider attacks traffic. The~recorded PCAPs are made available. Table~\ref{tab:attacks} summarises the traffic recorded per day. The~raw files of the CICIDS2017 dataset are pre-processed, as described in Section~\ref{sec:preprocessing}. The~full CICIDS2017 description and analysis is available in~\cite{panigrahi2018detailed} and~\cite{cicanalysis}.

\begin{table}[H]
    \centering
    \caption{CICIDS2017 attacks.} 
    \label{tab:attacks}
    \begin{tabular}{C{0.2\NetTableWidth}C{0.8\NetTableWidth}}
    \noalign{\hrule height 1.0pt}
        \rowcolor{gray!30}
         \textbf{Day} & \textbf{Traffic} \\
          \noalign{\hrule height 0.5pt}
         Monday & Benign \\
         \midrule
         Tuesday & SSH \& FTP Brute Force \\
         \midrule
         Wednesday & DoS/DDoS \& Heartbleed \\
         \midrule
         Thursday & Web Attack (Brute Force, XSS, Sql Injection) \& Infiltration \\
         \midrule
         Friday & Botnet, Portscan \& DDoS \\
         \bottomrule
    \end{tabular}
\end{table}

The second dataset is the NSL-KDD~\cite{RefWorks:doc:5b227bd4e4b0d1cffc0657b0}. NSL-KDD was released by the CIC in order to overcome the problems of the KDD Cup'99 dataset~\cite{RefWorks:doc:5d03bf33e4b0d6a9f11ef38c}. The~KDD Cup'99 dataset was the dataset of choice for evaluating more than 50\% of the past decade IDS~\cite{9108270}, followed by the NSL-KDD dataset, which was used for evaluating over 17\% of IDS. However, the~KDD Cup'99 has multiple drawbacks, as discussed thoroughly in~\cite{RefWorks:doc:5bbe0843e4b0e904d9061fbb}. These drawbacks include class imbalance and redundant records. Additionally, Siddique~{et al.}~\cite{siddique2019kdd} discussed the warnings provided to the UCI lab advising not to use KDD Cup'99 dataset in further IDS research.
Consequently, NSL-KDD fits for the evaluation purpose of this manuscript, as well as the comparison with relevant research.

The NSL-KDD dataset covers normal/benign traffic and four cyber-attack classes, namely, Denial of Service~(DoS), probing, Remote to Local~(R2L), and User to Root~(U2R). The~NSL-KDD dataset is available in two files `KDDTrain+.csv' and test file `KDDTest+.csv'.
Similar to the KDD Cup'99, the~NSL-KDD dataset is provided in comma separated value~(csv) feature files. Each instance is represented with its feature values alongside the class label.
The feature files undergo categorical features encoding to be appropriate for ML usage. The KDD Cup'99 and NSL-KDD datasets are analysed in~\cite{RefWorks:doc:5d03bf33e4b0d6a9f11ef38c}; furthermore, NSL-KDD is studied in~\cite{nslanalysis}.

\section{Methodology, Approach and Proposed Models}
\label{sec:proposed-models}

In this section, the~pre-processing of the datasets is discussed, followed by the explanation of the proposed, showing both the training and evaluation processes. Subsequently, Section~\ref{sec:results} details the evaluation and results. 

\subsection{CICIDS2017 Pre-Processing}
\label{sec:preprocessing}
The process that is involved in preparing the CICIDS2017 dataset for use is described as follows. Firstly,~`.pcap' files of the CICIDS2017 dataset are split based on the attack type and the timestamps provided by the dataset. This process results in a separate `.pcap' file for each attack class. Secondly, the~`.pcap' files are processed to generate bidirectional flows features. As highlighted by Rezaei
and Liu~\cite{rezaei2019deep}, with the advancement and complexity of networks and relying on encrypted traffic, features need to be suitable for both encrypted and unencrypted traffic analysis. The~authors also indicate that flow-based features are better-suited for modern IDS development. {Based on the analysis of recent IDSs by Aceto~et al.~\cite{ACETO2020306}, flow and bidirectional flow features are the most commonly used.} Thirdly,~features with high correlation are dropped in order to minimise model instability. 
Algorithm~\ref{alg:drop-features} describes the process of dropping highly correlated features. A threshold of `0.9' is used. Features with correlation less than the threshold are used for training. Finally, features are scaled using a Standard Scalar.
It is important to mention that only benign instances are used in selecting the features and scaling in order to ensure zero influence of attack instances. 

\begin{algorithm}[H]
    \caption{{Drop correlated }features}
    \label{alg:drop-features}
    \hspace*{\algorithmicindent} \textbf{Input:} Benign Data 2D Array, N,  Correlation Threshold \\
    \hspace*{\algorithmicindent} \textbf{Output:} Benign Data 2D Array, Dropped Columns
    \begin{algorithmic}[1]
    \setlength\baselineskip{17pt}
    \State ${correlation\_matrix} \leftarrow data.corr().abs()$ 
    \State $upper\_matrix \leftarrow correlation\_matrix[i, j]$ \hskip1cm $\{i, j \in N :i<= j\}$
    \State $dropped \leftarrow i \{i \in N: correlation\_matrix[i, ^*] > threshold\}$
    \State $data \leftarrow data.drop\_columns(dropped)$
    \State \textbf{return} $data$, $dropped$
    \end{algorithmic}
\end{algorithm}

As aforementioned, the~goal is to train models using benign traffic and evaluate their performance to detect attacks. Therefore, normal/benign traffic solely is used for training. The~normal instances are divided into 75\% for training and 25\% for testing/validation~\cite{split} by using sklearn train\_test\_split function with the shuffling option set to True (\url{https://scikit-learn.org/stable/modules/generated/sklearn.model_selection.train_test_split.html}). 
Furthermore, each of the attack classes then mimics a zero-day attack, thus assessing the ability of the model to detect its abnormality. Because the NSL-KDD dataset is split into training and testing, attacks in both files are used for evaluation. 

\subsection{Autoencoder-Based Model}
\label{sec:autoencoder}
The building block for the proposed Autoencoder is an Artificial Neural Network~(ANN). For~hyper-parameter optimisation, random search~\cite{bergstra2012random} is used in order to select the architecture of the network, number of epochs, and learning rate. Random search is known to converge faster than grid search to a semi-optimal set of parameters. It is also proved to be better than grid search when a small number of parameters are needed~\cite{liashchynskyi2019grid}. Finally, it limits the possibility of obtaining over-fitted parameters.  

Once the hyper-parameters are investigated, the~model is trained, as detailed in Algorithm~\ref{alg:training}. First,~the benign instances are split into 75\%:25\% for training and validation, respectively. Subsequently,~the~model is initialised using the optimal ANN architecture (number of layers and number of hidden neurons per layer). Finally, the~model is trained for $n$ number of epochs. The~loss and accuracy curves are observed in order to verify that the autoencoder convergence.

\begin{figure}[H]%
    \centering
    \includegraphics[width=0.7\linewidth]{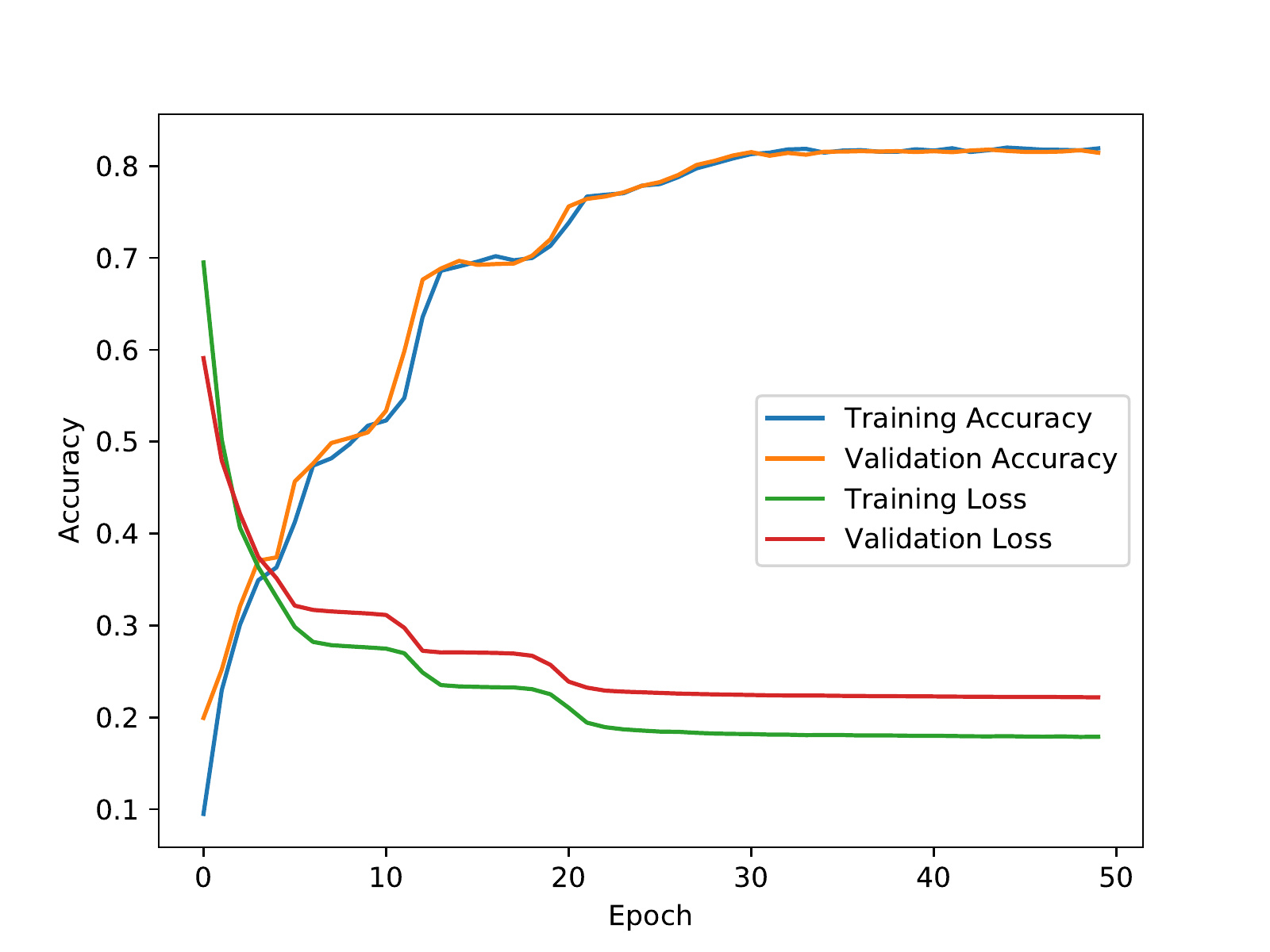}
    \caption{Autoencoder Convergence Curves}
    \label{fig:loss-curve}
\end{figure}

Once the model converges, as rendered in Figure~\ref{fig:loss-curve}, the~model is evaluated using Algorithm~\ref{alg:evaluation}. An~attack instance is flagged as a zero-day attack if the Mean Squared Error~(MSE) (reconstruction error) of the decoded ($x'$) and the original instance ($x$) is larger than a given threshold. For~the purpose of evaluation, multiple thresholds are assessed: 0.05, 0.1, 0.15. These thresholds are chosen based on the value that is chosen by the random search hyper-parameter optimisation.
The threshold plays an important role in deciding the value at which an instance is considered a zero-day attack, i.e.,~what MSE between $x'$ and $x$ is within the acceptable range.

\begin{algorithm}[H]
    \caption{Autoencoder Training}
    \label{alg:training}
    \hspace*{\algorithmicindent} \textbf{Input:} benign\_data, ANN\_architecture, regularisation\_value, num\_epochs \\
    \hspace*{\algorithmicindent} \textbf{Output:} Trained Autoencoder
    \begin{algorithmic}[1]
    \setlength\baselineskip{17pt}
    \State $training = 75\%\: i \in benign\_data$
    \State $testing = benign\_data\cap \overline{training}$

    \State $autoencoder \leftarrow build\_autoencoder($ $ANN\_Architecture, regularisation\_value)$
   \State $batch\_size \leftarrow 1024$
   \State $autoencoder.train(batch\_size, num\_epochs,$  $training, testing)$
    \State \textbf{return} $autoencoder$
    \end{algorithmic}
\end{algorithm}
\unskip
\begin{algorithm}[H]
    \caption{Evaluation}
    \label{alg:evaluation}
    \hspace*{\algorithmicindent} \textbf{Input:}  Trained Autoencoder, attack, thresholds \\
    \hspace*{\algorithmicindent} \textbf{Output:} Detection accuracies
    \begin{algorithmic}[1]
    \setlength\baselineskip{17pt}
    \State $detection\_accuracies \leftarrow \{\}$
    \State $predictions \leftarrow model.predict(attack)$
    \For{$th \in thresholds$}
        \State $accuracy \leftarrow$ $(mse(predictions, attack) > th)/len(attack) $
        \State $detection\_accuracies.add(threshold, accuracy)$
    \EndFor
    \State \textbf{return} $detection\_accuracies$
    \end{algorithmic}
\end{algorithm}
\unskip

\subsection{One-Class SVM based Model}
One-Class SVM is trained using the benign instances. In order to train the One-Class SVM, a `$\nu$' value was specified. As defined by Chen~{et al.},~``$\nu \in$ [0, 1], which is the lower and upper bound on the number of examples that are support vectors and that lie on the wrong side of the hyperplane, respectively.''~\cite{chen2005tutorial}.
The $\nu$ default value is 0.5, which~includes 50\% of the training sample in the hyperplane. However, for the purpose of this experiment, multiple $\nu$ values were chosen (0.2, 0.15,~0.1). These values were used in otder to evaluate and assess the autoencoder performance.

Algorithm~\ref{alg:svm-training} shows the process of training the One-Class SVM mode. Similar to the model that is discussed in Section~\ref{sec:autoencoder}, 75\% of the benign samples are used to fit the One-Class SVM model. 
Unlike~the Autoencoder model, where the evaluation relies on a threshold, a One-Class SVM trained model outputs a binary value \{0,1\}. The~output represents whether an instance belongs to the class to which the SVM is fit. Hence, each attack is evaluated based on how many instances are predicted with a `0' SVM output.

\begin{algorithm}[H]
    \caption{One-Class SVM Model}
    \label{alg:svm-training}
    \hspace*{\algorithmicindent} \textbf{Input:} benign\_data, nu\_value \\
    \hspace*{\algorithmicindent} \textbf{Output:} Trained SVM
    \begin{algorithmic}[1]
    \setlength\baselineskip{17pt}
    \State $training = 75\%\: i \in benign\_data$
    \State $testing = benign\_data\cap \overline{training}$

    \State $oneclasssvm \leftarrow OneClassSVM($ $nu\_value, `rbf')$
    \State $oneclasssvm.fit(training)$
    \State \textbf{return} $oneclasssvm$
    \end{algorithmic}
\end{algorithm}

\section{Experimental Results \label{sec:results}}
\unskip
\subsection{CICIDS2017 Autoencoder Results \label{sec:cic-ae-results}}

As mentioned, 75\% of the benign instances is used to train the Autoencoder. 
The autoencoder optimised architecture for the CICIDS2017 dataset is comprised from an ANN network with 18 neurons in both the input and the output layers and 3 hidden layers with 15, 9, 15 neurons respectively. The~optimal batch size is 1024. Other optimised parameters include mean square error loss, L2~regularisation of 0.0001 and for 50 epochs. 

\begin{figure}[H]
    \centering
    \includegraphics[width=1\textwidth, trim={0 4cm 1cm 0},clip]{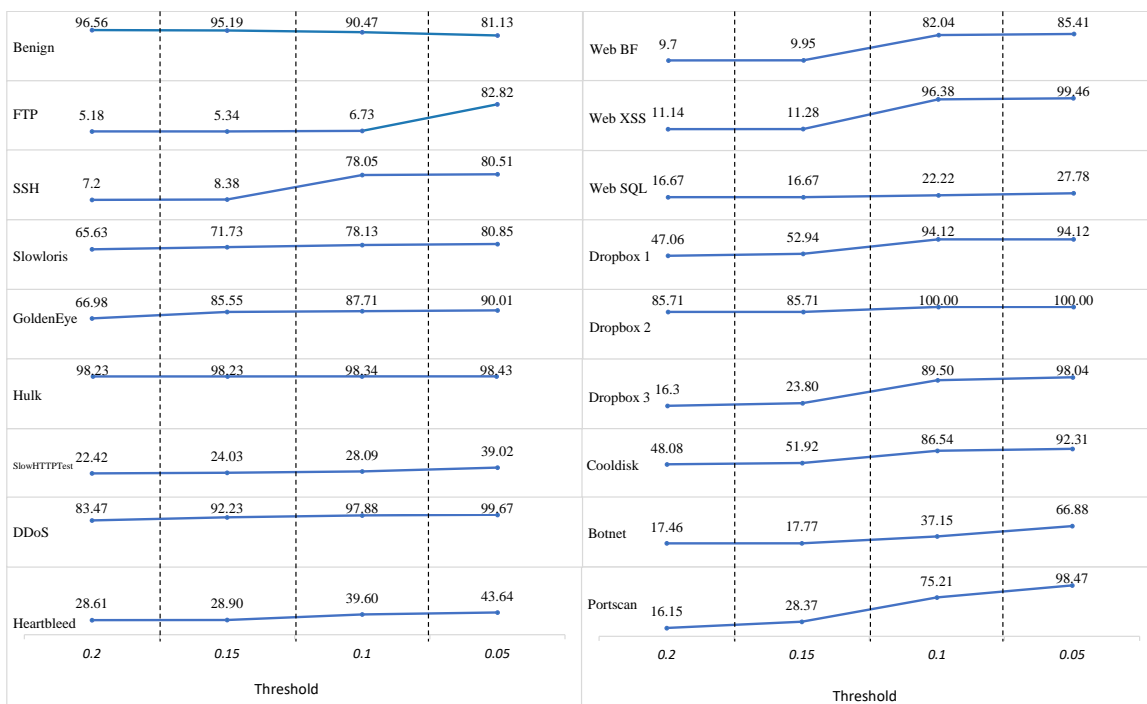}
    \caption{CICIDS2017 Autoencoder Detection Results Summary Per Class}
    \label{fig:cic-AE-summary}
\end{figure}

Figure~\ref{fig:cic-AE-summary} summarises the autoencoder accuracy of all CICIDS2017 classes. It is crucial to note that accuracy is defined differently for benign. Unlike attacks, for benign class, the~accuracy represents the rate of instances not classified as zero-day (i.e.,~benign) which reflects the specificity and for the attack classes it represents the recall.
\vspace{-6pt}

By observing Figure~\ref{fig:cic-AE-summary}, benign accuracy is 95.19\%, 90.47\% and 81.13\% for a threshold of 0.15, 0.1 and 0.05 respectively. Moreover, for the different attack detection accuracy, it is observed that there are three categories. Firstly, attacks that are very different from benign (for example, Hulk and DDoS), the~detection accuracy is high regardless the threshold [92--99\%]. Secondly, classes that are slightly different from benign (for example, SSH Brute-force and Port scanning), an accuracy rise is observed for lower thresholds. This emphasise the threshold's role. Thirdly, classes that are not distinguishable from benign traffic, they are detected but with a lower accuracy (for example, Botnet, SQL Injection and DoS-SlowHTTPTest).

By observing Figure~\ref{fig:cic-AE-summary}, different categories can be seen, (a) classes with a stable detection accuracy (i.e.,~ line), and (b) classes with a prompt rise in detection accuracy in the right-most slice (0.05 threshold). Finally, the~benign accuracy (top left) falls within an acceptable range with different thresholds.

\subsection{CICIDS2017 One-Class SVM Results}

Table~\ref{tab:cicids-svm-results} summarises the One-Class SVM results. By observing the One-Class SVM results, two~assertions are identified, (a) the  detection accuracy is not affected significantly by changing $\nu$ value, and (b) the classes with high detection accuracy in the Autoencoder results (Figure~\ref{fig:cic-AE-summary} 
 are also detected by the One-Class SVM; however, the~One-Class SVM fails to detect the two other categories (rise in detection accuracy with small thresholds and low detection accuracy). This is due to the limitations of the One-Class SVM algorithm which attempts to fit a spherical hyperplane to separate benign class from other classes, however, classes that fall into this hyperplane will always be classified as benign/normal.

\begin{figure}[H]
    \centering
    \begin{subfigure}[t]{0.75\textwidth}
        \centering
        \includegraphics[width=0.9\linewidth, trim={1.5cm 5cm 1.5cm 4cm},clip]{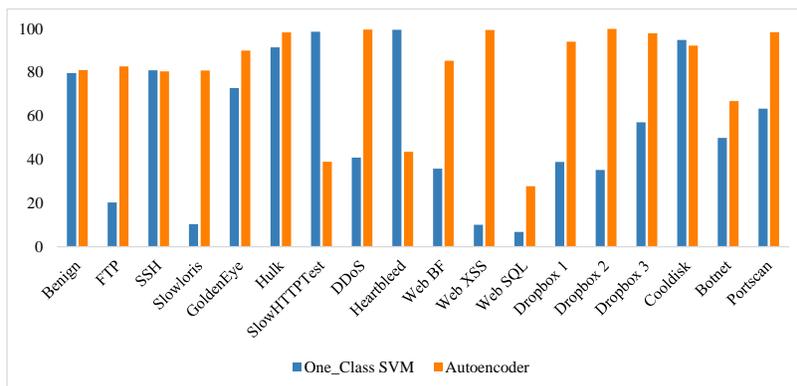}
        \caption{SVM ($\nu = 0.2 $), AE (Threshold = 0.05)}
    \end{subfigure}%

      \begin{subfigure}[t]{0.75\textwidth}
        \centering
        \includegraphics[width=0.97\linewidth, trim={1.5cm 5cm 1.5cm 4cm},clip]{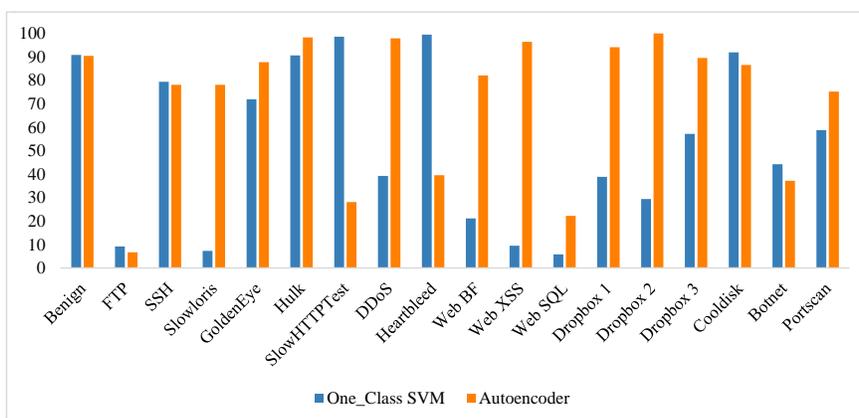}
        \caption{SVM ($\nu = 0.09 $), AE (Threshold = 0.1)}
    \end{subfigure}
    \vspace{2mm}
    \caption{CICIDS2017 Autoencoder, One-Class SVM Comparison.}
    \label{fig:cic-ae-svm}
\end{figure}

This can further be visualised in Figure~\ref{fig:cic-ae-svm}. One-Class SVM is well suited for flagging recognisable zero-day attacks. However, autoencoders are better suited for complex zero-day attacks as the performance rank is significantly higher. Furthermore, Figure~\ref{fig:cic-ae-svm} shows a class by class comparison of the performance of autoencoder versus One-Class SVM. Figure~\ref{fig:cic-ae-svm}a plots the results using One-Class SVM $\nu = 0.2$ and autoencoder threshold of 0.05, while Figure~\ref{fig:cic-ae-svm}b plots the results using One-Class SVM $\nu = 0.09$ and autoencoder threshold of 0.1. 

\begin{table}[H]
    \centering
    \caption{CICIDS2017 One-Class Support Vector Machine (SVM) Results.}
    \label{tab:cicids-svm-results}
    \begin{tabular}{C{0.4\NetTableWidth}C{0.2\NetTableWidth}C{0.2\NetTableWidth}C{0.2\NetTableWidth}}
        \noalign{\hrule height 1.0pt}
        \rowcolor{gray!30}
        \textbf{Class} & \multicolumn{3}{c}{\textbf{Accuracy}} \\ \noalign{\hrule height 0.5pt}
        \rowcolor{gray!30}
        \boldmath{$\nu$} & \textbf{0.2} & \textbf{0.15} & \textbf{0.1} \\ \noalign{\hrule height 0.5pt}
        Benign (Validation) & 89.81\% & 84.84\% & 79.71\% \\ \midrule 
        FTP Bruteforce & 10.19\% & 15.16\% & 20.29\% \\ \midrule 
        SSH Bruteforce & 79.51\% & 80.26\% & 80.95\% \\ \midrule 
        DoS (Slowloris) & 7.66\% & 8.38\% & 10.37\% \\ \midrule 
        DoS (GoldenEye) & 71.87\% & 72.39\% & 72.85\% \\ \midrule 
        DoS (Hulk) & 90.69\% & 91.35\% & 91.55\% \\ \midrule 
        DoS (Slowhttps) & 98.59\% & 98.66\% & 98.71\% \\ \midrule 
        DDoS & 39.35\% & 39.94\% & 40.96\% \\ \midrule
        Heartbleed & 99.49\% & 99.54\% & 99.58\% \\ \midrule 
        Web BF & 21.1\% & 23.41\% & 35.84\% \\ \midrule 
        Web XSS & 9.58\% & 9.76\% & 10.13\% \\ \midrule 
        Web SQL & 5.77\% & 6.31\% & 6.85\% \\ \midrule 
        Infiltration - Dropbox 1 & 38.89\% & 38.89\% & 38.89\% \\ \midrule 
        Infiltration - Dropbox 2 & 29.41\% & 35.29\% & 35.29\% \\ \midrule 
        Infiltration - Dropbox 3 & 57.14\% & 57.14\% & 57.14\% \\ \midrule 
        Infiltration - Cooldisk & 92.15\% & 93.8\% & 94.91\% \\ \midrule 
        Botnet & 44.23\% & 46.15\% & 50\% \\ \midrule 
        PortScan & 59.27\% & 60.04\% & 63.43\% \\ \bottomrule 
        \end{tabular}
    \end{table}
\unskip


\subsection{NSL-KDD Results}
The autoencoder optimised architecture for the NSL-KDD dataset is comprised from an ANN network with 122 neurons in both the input and output layers and three hidden layers with 100, 60, 100 neurons, respectively. The~optimal batch size is 1024. Other optimised parameters include mean absolute error loss, L2 regularisation of 0.001, and for 50 epochs.


Table~\ref{tab:nsl-kdd-ae-results} shows the autoencoder results for the NSL-KDD dataset. As aforementioned, attacks in both the KDDTrain+ and KDDTest+ files are used in order to evaluate the model. 
Similar to the results that are discussed in Section~\ref{sec:cic-ae-results}, the~trade-off between the threshold choice and the true negative rate is~observed. 

\begin{table}[H]
    \centering
    \caption{NSL-KDD Autoencoder Results}
    \label{tab:nsl-kdd-ae-results}
    \begin{tabular}{C{0.4\NetTableWidth}C{0.2\NetTableWidth}C{0.2\NetTableWidth}C{0.2\NetTableWidth}}
        \noalign{\hrule height 1.0pt}
        \rowcolor{gray!30}
        \textbf{Class} & \multicolumn{3}{c}{\textbf{Accuracy}} \\ \noalign{\hrule height 0.5pt}
        \rowcolor{gray!30}
        \boldmath{$Threshold$} & \textbf{0.3} & \textbf{0.25} & \textbf{0.2} \\ \noalign{\hrule height 0.5pt}

        \rowcolor{gray!30}
        \multicolumn{4}{c}{KDDTrain+.csv} \\ \noalign{\hrule height 0.5pt}
        Normal (Validation) & 79.09\% & 77.80\% & 72.78\%\\ \midrule 
        DoS & 98.15\% & 98.16\% & 98.17\% \\ \midrule 
        Probe & 99.89\% & 99.94\% & 99.94\% \\ \midrule 
        R2L & 83.12\% & 96.48\% & 96.48\% \\ \midrule 
        U2R & 84.62\% & 100\% & 100\% \\ \noalign{\hrule height 0.5pt}
        \rowcolor{gray!30}
        \multicolumn{4}{c}{KDDTest+.csv} \\ \noalign{\hrule height 0.5pt}
        Normal & 84.82\% & 84.42\% & 80.94\% \\ \midrule 
        DoS & 94.67\% & 94.67\% & 94.76\% \\ \midrule 
        Probe & 100\% & 100\% & 100\%  \\ \midrule 
        R2L & 95.95\% & 96.5\% & 97\%\\ \midrule 
        U2R & 83.78\% & 89.19\% & 100\% \\ \midrule 
    \end{tabular}
    \end{table}

Furthermore, when compared to the most recent available autoencoder implementation for detecting zero-day attacks in the literature~\cite{gharib2019autoids}, the~autoencoder that is proposed in this manuscript largely outperforms the performances of~\cite{gharib2019autoids}. The~work proposed by Gharib~{et al.}~\cite{gharib2019autoids} used a hybrid two stage autoencoder to detect normal and abnormal traffic. Training on KDDTrain+ file and testing on KDDTest+, the~overall accuracy of their proposed model is 90.17\%, whereas the proposed autoencoder in this manuscript the overall accuracy is 91.84\%, 92.96\%, and 91.84\% using a threshold of 0.3, 0.25, and 0.2, respectively. It is important to highlight that Gharib~{et al.}~\cite{gharib2019autoids} do not mention details regarding how they define anomalies or zero-day attacks or the classes they use in the testing process.
Moreover, as summarised in Table~\ref{tab:nsl-compare}, it is shown that the proposed approach in this manuscript outperforms the Denoising Autoencoder that is proposed in~\cite{aygun2017network}, specifically with the KDDTest+ instances with the authors accuracy is capped by 88\%, while this manuscript reaches 94\%.

\begin{table}[H]
    \centering
    \caption{NSL-KDD Performance Comparison with Recent Literature. (Highest accuracy in \textbf{Bold})}
    \begin{tabular}{C{0.05\NetTableWidth}C{0.12\NetTableWidth}C{0.2\NetTableWidth}C{0.22\NetTableWidth}C{0.16\NetTableWidth}C{0.16\NetTableWidth}}
        \noalign{\hrule height 1.0pt}
         \rowcolor{gray!30}
        \textbf{Year} & \textbf{Reference} & \textbf{Approach} & \textbf{Train:Test \% of KDDTrain+} & \textbf{KDDTrain+ Accuracy} & \textbf{KDDTest+ Accuracy} \\ \noalign{\hrule height 0.5pt}
        \multicolumn{2}{c}{\textbf{This paper}} 
       & AE th = 0.3 \newline \textbf{AE th = 0.25} \newline AE th = 0.2 & 75 :  25 & 88.97\% \newline \textbf{94.48}\% \newline 93.47\% & 91.84\% \newline 92.96\%	\newline \textbf{94.54}\% \\ \midrule

        2019 & \cite{gharib2019autoids} & 2 AEs & - & - & 90.17\% \\ \midrule 
        2017 & \cite{aygun2017network} & AE \newline Denoising AE & 80 : 20 &  93.62\% \newline 94.35\% &  88.28\% \newline 88.65\%  \\ \midrule
    \end{tabular}
    \label{tab:nsl-compare}
\end{table}

Table~\ref{tab:nsl-kdd-svm-results} summarises the NSL-KDD One-Class SVM results. The~results show a similar detection trend. This is due to the limited number and variance of attacks that are covered by the NSL-KDD dataset.

\begin{table}[H]
    \centering
    \caption{NSL-KDD One-Class SVM Results}
    \label{tab:nsl-kdd-svm-results}
    \begin{tabular}{C{0.4\NetTableWidth}C{0.2\NetTableWidth}C{0.2\NetTableWidth}C{0.2\NetTableWidth}}
        \noalign{\hrule height 1.0pt}
        \rowcolor{gray!30}
        \textbf{Class} & \multicolumn{3}{c}{\textbf{Accuracy}} \\ \noalign{\hrule height 0.5pt}
        \rowcolor{gray!30}
       \boldmath{ $\nu$} & \textbf{0.2} & \textbf{0.15} & \textbf{0.1} \\ \noalign{\hrule height 0.5pt}
        \rowcolor{gray!30}
        \multicolumn{4}{c}{KDDTrain+.csv} \\ \noalign{\hrule height 0.5pt}
        Normal (Validation) & 89.9\% & 85.14\% & 80.54\% \\ \midrule 
        DoS & 98.13\% & 98.14\% & 98.14\% \\ \midrule 
        Probe & 97.74\% & 98.77\% & 99.52\% \\ \midrule 
        R2L & 49.35\% & 52.26\% & 81.71\% \\ \midrule 
        U2R & 78.85\% & 80.77\% & 82.69\% \\ \noalign{\hrule height 0.5pt}
        \rowcolor{gray!30}
        \multicolumn{4}{c}{KDDTest+.csv} \\ \noalign{\hrule height 0.5pt}
        Normal & 88.12\% & 86.02\% & 84.72\% \\ \midrule 
        DoS & 94.67\% & 94.67\% & 94.69\% \\ \midrule 
        Probe & 99.55\% & 99.91\% & 100\% \\ \midrule 
        R2L & 80.17\% & 82.22\% & 90.31\% \\ \midrule 
        U2R & 78.38\% & 78.38\% & 83.78\% \\ \midrule 

    \end{tabular}
    
\end{table}

\newpage

\section{Conclusions and Future Work}
\label{sec:conclusion}

The work that is presented in this manuscript proposes a new outlier-based zero-day cyber-attacks detection. The~main goal was to develop an intelligent IDS model that is capable of detecting zero-day cyber-attacks with a high detection accuracy while overcoming the limitations of currently available IDS. This manuscript purposes and evaluates an autoencoder model to detect zero-day attacks. 
The~idea is inspired by the encoding-decoding capability of autoencoders. 

The results show high detection accuracy for the autoencoder model for both the CICIDS2017 and the NSL-KDD. The~CICIDS2017 zero-day detection accuracy reaches 90.01\%, 98.43\%, 98.47\%, and 99.67\%  for DoS (GoldenEye), DoS (Hulk), Port scanning, and DDoS attacks. Moreover, the~NSL-KDD detection accuracy reached 92.96\%, which~outperforms the only available zero-day autoencoder-based detection manuscript~\cite{gharib2019autoids}.

Furthermore, the~autoencoder model is compared to an unsupervised outlier-based ML technique; One-Class SVM. One-Class SVM is a prominent unsupervised ML technique that detects outliers. 
The~one-class SVM mode presents its effectiveness in detecting zero-day attacks for NSL-KDD datasets and the distinctive ones from the CICIDS2017 dataset. When compared to One-Class SVM, autoencoder demonstrates its surpassing detection accuracy. Furthermore, both of the models demonstrate low miss rate (false-positives).
Future work involves evaluating the proposed models with datasets that cover special purpose network IDS (e.g., IoT and Critical Infrastructure networks), which~will comprise insights into adapting the proposed models, as well as proposing and adapting other ML techniques to use for zero-day attack detection.
The source code for building and evaluating the proposed models will be made available through an open-source GitHub repository.

\vspace{6pt} 



\authorcontributions{Conceptualization, H.H., R.A. and X.B.; Formal analysis, H.H., R.A. and J.-N.C.; Investigation, H.H.; Methodology, H.H., R.A., C.T. and X.B.; Project administration, X.B.; Software, H.H.; Supervision, E.B. and X.B.; Validation, R.A., C.T., E.B. and X.B.; Writing---original draft, H.H.; Writing---review \& editing, R.A., J.-N.C., E.B. and X.B. All authors have read and agreed to the published version of the manuscript.}

\funding{This research received no external funding.}


\conflictsofinterest{The authors declare no conflict of interest.}
\reftitle{References}






\begin{thebibliography}{999}

\bibitem[Kaloudi and Li(2020)]{10.1145/3372823}
Kaloudi, N.; Li, J.
\newblock The AI-Based Cyber Threat Landscape: A Survey.
\newblock {\em ACM Comput. Surv.} {\bf 2020}, {\em 53},
\newblock
  doi:{\changeurlcolor{black}\href{https://doi.org/10.1145/3372823}{\detokenize{10.1145/3372823}}}.

\bibitem[{Hindy} \em{et~al.}(2018){Hindy}, {Hodo}, {Bayne}, {Seeam},
  {Atkinson}, and {Bellekens}]{8551386}
{Hindy}, H.; {Hodo}, E.; {Bayne}, E.; {Seeam}, A.; {Atkinson}, R.; {Bellekens},
  X.
\newblock A Taxonomy of Malicious Traffic for Intrusion Detection Systems.
\newblock In~Proceedings of~the 2018 International Conference On Cyber Situational Awareness, Data Analytics and Assessment {(Cyber SA)}, Glasgow, Scotland, UK, 11--12 June 2018; pp.~1--4. 

\bibitem[Khraisat \em{et~al.}(2019)Khraisat, Gondal, Vamplew, and
  Kamruzzaman]{Khraisat2019}
Khraisat, A.; Gondal, I.; Vamplew, P.; Kamruzzaman, J.
\newblock Survey of Intrusion Detection Systems: tTechniques, Datasets and
  Challenges.
\newblock {\em Cybersecurity} {\bf 2019}, {\em 2},~20,
\newblock
  doi:{\changeurlcolor{black}\href{https://doi.org/10.1186/s42400-019-0038-7}{\detokenize{10.1186/s42400-019-0038-7}}}.

\bibitem[{Hindy} \em{et~al.}(2018){Hindy}, {Brosset}, {Bayne}, {Seeam},
  {Tachtatzis}, {Atkinson}, and {Bellekens}]{HananHindy2018}
{Hindy}, H.; {Brosset}, D.; {Bayne}, E.; {Seeam}, A.; {Tachtatzis}, C.;
  {Atkinson}, R.; {Bellekens}, X.
\newblock A Taxonomy and Survey of Intrusion Detection System Design
  Techniques, Network Threats and Datasets.
\newblock {\em arXiv} {\bf 2018}, arXiv:1806.03517.

\bibitem[Chapman(2016)]{CHAPMAN20161}
Chapman, C.
\newblock Chapter 1---Introduction to Practical Security and Performance
  Testing. In~{\em Network Performance and Security}; Chapman, C., Ed.;
  Syngress: Boston, MA, USA, 2016; pp.~1--14,
\newblock
  doi:{\changeurlcolor{black}\href{https://doi.org/https://doi.org/10.1016/B978-0-12-803584-9.00001-9}{\detokenize{10.1016/B978-0-12-803584-9.00001-9}}}.

\bibitem[Bilge and Dumitra\c{s}(2012)]{10.1145/2382196.2382284}
Bilge, L.; Dumitra\c{s}, T.
\newblock Before We Knew It: An Empirical Study of Zero-Day Attacks in the Real
  World.
\newblock  In~Proceedings of~the 2012 ACM Conference on Computer and
  Communications {Security} (CCS ’12), Raleigh, NC, USA, 16--18 October 2012; pp.~833--844,
\newblock 
  doi:{\changeurlcolor{black}\href{https://doi.org/10.1145/2382196.2382284}{\detokenize{10.1145/2382196.2382284}}}.

\bibitem[Nguyen and Reddi(2019)]{nguyen2019deep}
Nguyen, T.T.; Reddi, V.J.
\newblock Deep Reinforcement Learning for Cyber Security.
\newblock {\em arXiv} {\bf 2019},  arXiv:1906.05799.

\bibitem[Metrick \em{et~al.}(2020)Metrick, Najafi, and
  Semrau]{ZeroDayE16:online}
Metrick, K.; Najafi, P.; Semrau, J.
\newblock \emph{Zero-Day Exploitation Increasingly Demonstrates Access to Money,
  Rather than Skill---Intelligence for Vulnerability Management}; Part One; {FireEye Inc.}: Milpitas, CA, USA, 2020.
\newblock  

\bibitem[Cisco(2017)]{cisco2017}
Cisco.
\newblock Cisco 2017 Annual Cyber security Report. 2017.
\newblock  Available online: \url{https://www.grouppbs.com/wp-content/uploads/2017/02/Cisco_2017_ACR_PDF.pdf} (accessed on 20 July 2020).

\bibitem[{Ficke} \em{et~al.}(2019){Ficke}, {Schweitzer}, {Bateman}, and
  {Xu}]{9020860}
{Ficke}, E.; {Schweitzer}, K.M.; {Bateman}, R.M.; {Xu}, S.
\newblock Analyzing Root Causes of Intrusion Detection False-Negatives:
  Methodology and Case Study.
\newblock  In~Proceedings of~the 2019 IEEE Military Communications Conference (MILCOM), Norfolk, VA, USA, 12--14 November {2019}; pp.~1--6.

\bibitem[Sharma \em{et~al.}(2018)Sharma, Kim, Kwon, You, Lee, and
  Yim]{sharma2018framework}
Sharma, V.; Kim, J.; Kwon, S.; You, I.; Lee, K.; Yim, K.
\newblock A Framework for Mitigating Zero-Day Attacks in IoT.
\newblock {\em arXiv} {\bf 2018},  arXiv:1804.05549.

\bibitem[Sun \em{et~al.}(2018)Sun, Dai, Liu, Singhal, and
  Yen]{RefWorks:doc:5c812934e4b075899776c5dd}
Sun, X.; Dai, J.; Liu, P.; Singhal, A.; Yen, J.
\newblock Using Bayesian Networks for Probabilistic Identification of Zero-Day
  Attack Paths.
\newblock {\em IEEE Trans. Inf. Forensics Secur.} {\bf 2018}, {\em 13},~2506--2521,
\newblock doi:{\changeurlcolor{black}\href{https://doi.org/10.1109/TIFS.2018.2821095}{\detokenize{10.1109/TIFS.2018.2821095}}}.

\bibitem[Zhou and Pezaros(2019)]{zhou2019evaluation}
Zhou, Q.; Pezaros, D.
\newblock Evaluation of Machine Learning Classifiers for Zero-Day Intrusion
  Detection--An Analysis on CIC-AWS-2018 dataset.
\newblock {\em arXiv} {\bf 2019}, arXiv:1905.03685.

\bibitem[Zhao \em{et~al.}(2019)Zhao, Shetty, Pan, Kamhoua, and
  Kwiat]{zhao2019transfer}
Zhao, J.; Shetty, S.; Pan, J.W.; Kamhoua, C.; Kwiat, K.
\newblock Transfer Learning for Detecting Unknown Network Attacks.
\newblock {\em EURASIP J. Inf. Secur.} {\bf 2019}, {\em 2019},~1.

\bibitem[Sameera and Shashi(2020)]{SAMEERA2020}
Sameera, N.; Shashi, M.
\newblock Deep Transductive Transfer Learning Framework for Zero-Day Attack
  Detection.
\newblock {\em ICT Express} {\bf 2020},
\newblock
  doi:{\changeurlcolor{black}\href{https://doi.org/https://doi.org/10.1016/j.icte.2020.03.003}{\detokenize{10.1016/j.icte.2020.03.003}}}.

\bibitem[Abri \em{et~al.}(2019)Abri, Siami-Namini, Khanghah, Soltani, and
  Namin]{RefWorks:doc:5dde4cdee4b0518003692b09}
Abri, F.; Siami-Namini, S.; Khanghah, M.A.; Soltani, F.M.; Namin, A.S.
\newblock The Performance of Machine and Deep Learning Classifiers in Detecting
  Zero-Day Vulnerabilities.
\newblock {\em arXiv} {\bf 2019}, arXiv:1911.09586.

\bibitem[Kim \em{et~al.}(2018)Kim, Bu, and
  Cho]{RefWorks:doc:5ddeb66fe4b0227c51200fde}
Kim, J.Y.; Bu, S.J.; Cho, S.B.
\newblock Zero-day Malware Detection using Transferred Generative Adversarial
  Networks based on Deep Autoencoders. \emph{Information Sciences} \textbf{2018}, \emph{460–461}, 83--102, 
\newblock doi:{\changeurlcolor{black}\href{https://doi.org/https://doi.org/10.1016/j.ins.2018.04.092}{\detokenize{10.1016/j.ins.2018.04.092}}}.

\bibitem[Rumelhart \em{et~al.}(1985)Rumelhart, Hinton, and
  Williams]{RefWorks:doc:5dd2bfdfe4b0a03ea81aa826}
Rumelhart, D.E.; Hinton, G.E.; Williams, R.J.
\newblock \emph{Learning Internal Representations by Error Propagation};
\newblock Technical Report; California Univ San Diego La Jolla Inst for
  Cognitive Science: San Diego, CA, USA, 1985.

\bibitem[Stewart(2019)]{Comprehe85}
Stewart, M.
\newblock Comprehensive Introduction to Autoencoders. 2019.
\newblock  Available online: \url{https://towardsdatascience.com/generating-images-with-autoencoders-77fd3a8dd368} (accessed on 21 July 2020).

\bibitem[Goodfellow \em{et~al.}(2016)Goodfellow, Bengio, and
  Courville]{RefWorks:doc:5dd2b90be4b0a03ea81aa557}
Goodfellow, I.; Bengio, Y.; Courville, A.
\newblock {\em Deep Learning}; {MIT Press}: Cambridge, MA, USA, 2016.

\bibitem[Barber(2014)]{barber2014implicit}
Barber, D.
\newblock \emph{Implicit Representation Networks};
\newblock Technical Report; Department of Computer Science, 
 University College London: London, UK, 2014.

\bibitem[Hinton and Salakhutdinov(2006)]{RefWorks:doc:5dd2b992e4b0ff5398154ceb}
Hinton, G.E.; Salakhutdinov, R.R.
\newblock Reducing the Dimensionality of Data with Neural Networks.
\newblock {\em Science} {\bf 2006}, {\em 313},~504--507.

\bibitem[Zabalza \em{et~al.}(2016)Zabalza, Ren, Zheng, Zhao, Qing, Yang, Du,
  and Marshall]{RefWorks:doc:5dd2b944e4b0a03ea81aa581}
Zabalza, J.; Ren, J.; Zheng, J.; Zhao, H.; Qing, C.; Yang, Z.; Du, P.;
  Marshall, S.
\newblock Novel Segmented Stacked Autoencoder for Effective Dimensionality
  Reduction and Feature Extraction in Hyperspectral Imaging.
\newblock {\em Neurocomputing} {\bf 2016}, {\em 185},~1--10.

\bibitem[Liou \em{et~al.}(2014)Liou, Cheng, Liou, and
  Liou]{RefWorks:doc:5dd2ba6ee4b039d5609d627e}
Liou, C.Y.; Cheng, W.C.; Liou, J.W.; Liou, D.R.
\newblock Autoencoder for Words. \emph{Neurocomputing}  \textbf{2014}, \emph{139}, 84--96.
  doi:{\changeurlcolor{black}\href{https://doi.org/https://doi.org/10.1016/j.neucom.2013.09.055}{\detokenize{10.1016/j.neucom.2013.09.055}}}.

\bibitem[Theis \em{et~al.}(2017)Theis, Shi, Cunningham, and
  Huszár]{RefWorks:doc:5dd2bd88e4b0caca97c355e0}
Theis, L.; Shi, W.; Cunningham, A.; Huszár, F.
\newblock Lossy Image Compression with Compressive Autoencoders.
\newblock {\em arXiv} {\bf 2017}, arXiv:1703.00395.

\bibitem[Zhou and Paffenroth(2017)]{RefWorks:doc:5dd2bc1ae4b0c0c541dadb08}
Zhou, C.; Paffenroth, R.C.
\newblock Anomaly Detection with Robust Deep Autoencoders.
\newblock In~Proceedings of~the 23rd ACM SIGKDD International Conference on
  Knowledge Discovery and Data Mining, Halifax, NS, Canada, 13--17 August {2017;} pp.~665--674.

\bibitem[Creswell and Bharath(2018)]{RefWorks:doc:5dd2b844e4b064aae78e03e6}
Creswell, A.; Bharath, A.A.
\newblock Denoising Adversarial Autoencoders.
\newblock {\em IEEE Trans. Neural Networks Learn. Syst.} {\bf 2018}, {\em 30},~968--984.

\bibitem[Cortes and Vapnik(1995)]{cortes1995support}
Cortes, C.; Vapnik, V.
\newblock Support-Vector Networks.
\newblock {\em Mach. Learn.} {\bf 1995}, {\em 20},~273--297.

\bibitem[Ng(2000)]{ng2000cs229}
Ng, A.
\newblock Part V: Support Vector Machines|CS229 Lecture Notes. 2000.

\bibitem[Zisserman(2015)]{Hilary2015}
Zisserman, A.
\newblock The SVM classifier|C19 Machine Learning. 2015.

\bibitem[Sch{\"o}lkopf \em{et~al.}(2000)Sch{\"o}lkopf, Williamson, Smola,
  Shawe-Taylor, and Platt]{scholkopf2000support}
Sch{\"o}lkopf, B.; Williamson, R.C.; Smola, A.J.; Shawe-Taylor, J.; Platt, J.C.
\newblock Support Vector Method for Novelty Detection.
\newblock  In \emph{Advances in Neural Information Processing Systems 12};  2000; pp.~582--588. MIT Press 

\bibitem[Tax and Duin(2004)]{tax2004support}
Tax, D.M.; Duin, R.P.
\newblock Support Vector Data Description.
\newblock {\em Mach. Learn.} {\bf 2004}, {\em 54},~45--66.

\bibitem[Wang \em{et~al.}(2018)Wang, Liu, Zhu, Porikli, and Yin]{WANG2018198}
Wang, S.; Liu, Q.; Zhu, E.; Porikli, F.; Yin, J.
\newblock Hyperparameter Selection of One-Class Support Vector Machine by
  Self-Adaptive Data Shifting.
\newblock {\em Pattern Recognit.} {\bf 2018}, {\em 74},~198--211,
\newblock
  doi:{\changeurlcolor{black}\href{https://doi.org/https://doi.org/10.1016/j.patcog.2017.09.012}{\detokenize{10.1016/j.patcog.2017.09.012}}}.

\bibitem[Hodo \em{et~al.}(2017)Hodo, Bellekens, Hamilton, Tachtatzis, and
  Atkinson]{hodo2017shallow}
Hodo, E.; Bellekens, X.; Hamilton, A.; Tachtatzis, C.; Atkinson, R.
\newblock Shallow and Deep Networks Intrusion Detection System: A Taxonomy and
  Survey.
\newblock {\em arXiv} {\bf 2017}, arXiv:1701.02145.

\bibitem[Hamed \em{et~al.}(2018)Hamed, Ernst, and
  Kremer]{RefWorks:doc:5cdd85f5e4b02cf23c27b848}
Hamed, T.; Ernst, J.B.; Kremer, S.C.
\newblock A Survey and Taxonomy of Classifiers of Intrusion Detection Systems.
  In~{\em Computer and Network Security Essentials}; Computer
  and Network Security Essentials; Daimi, K., Ed.; Springer International Publishing: Cham, Switzerland, 2018; pp.~21--39,
\newblock
  doi:{\changeurlcolor{black}\href{https://doi.org/10.1007/978-3-319-58424-9\_2}{\detokenize{10.1007/978-3-319-58424-9\_2}}}.

\bibitem[{Hindy} \em{et~al.}(2020){Hindy}, {Brosset}, {Bayne}, {Seeam},
  {Tachtatzis}, {Atkinson}, and {Bellekens}]{9108270}
{Hindy}, H.; {Brosset}, D.; {Bayne}, E.; {Seeam}, A.K.; {Tachtatzis}, C.;
  {Atkinson}, R.; {Bellekens}, X.
\newblock A Taxonomy of Network Threats and the Effect of Current Datasets on
  Intrusion Detection Systems.
\newblock {\em IEEE Access} {\bf 2020}, {\em 8},~104650--104675.

\bibitem[Buczak and Guven(2016)]{7307098}
Buczak, A.L.; Guven, E.
\newblock A Survey of Data Mining and Machine Learning Methods for Cyber
  Security Intrusion Detection.
\newblock {\em IEEE Commun. Surv. Tutor.} {\bf 2016}, {\em
  18},~1153--1176,
\newblock
  doi:{\changeurlcolor{black}\href{https://doi.org/10.1109/COMST.2015.2494502}{\detokenize{10.1109/COMST.2015.2494502}}}.

\bibitem[Aceto \em{et~al.}(2020)Aceto, Ciuonzo, Montieri, and
  Pescapé]{ACETO2020306}
Aceto, G.; Ciuonzo, D.; Montieri, A.; Pescapé, A.
\newblock {Toward effective mobile encrypted traffic classification through
  deep learning}.
\newblock {\em Neurocomputing} {\bf 2020}, {\em 409},~306--315,
\newblock
  doi:{\changeurlcolor{black}\href{https://doi.org/https://doi.org/10.1016/j.neucom.2020.05.036}{\detokenize{10.1016/j.neucom.2020.05.036}}}.

\bibitem[Rashid(2019)]{Encrypti14}
Rashid, F.Y.
\newblock Encryption, Privacy in the Internet Trends Report|Decipher. 2019.
\newblock  Available online: \url{https://duo.com/decipher/encryption-privacy-in-the-internet-trends-report} (accessed on 14 September 2020).

\bibitem[{Kunang} \em{et~al.}(2018){Kunang}, {Nurmaini}, {Stiawan}, {Zarkasi},
  {Firdaus}, and {Jasmir}]{8605181}
{Kunang}, Y.N.; {Nurmaini}, S.; {Stiawan}, D.; {Zarkasi}, A.; {Firdaus}.;
  {Jasmir}.
\newblock Automatic Features Extraction Using Autoencoder in Intrusion
  Detection System.
\newblock In~Proceedings of~the 2018 International Conference on Electrical Engineering and Computer  Science (ICECOS), Pangkal Pinang, Indonesia, 2--4 October {2018;} pp.~219--224.

\bibitem[Kherlenchimeg and Nakaya(2018)]{RefWorks:doc:5f3a7005e4b07d1d33a62c96}
Kherlenchimeg, Z.; Nakaya, N.
\newblock Network Intrusion Classifier Using Autoencoder with Recurrent Neural
  Network.
\newblock  In~Proceedings of~the Fourth International Conference on Electronics
  and Software Science (ICESS2018), Takamatsu, Japan, 5--7 November 2018; pp.~94--100.

\bibitem[Shaikh and Shashikala(2019)]{shaikh2019autoencoder}
Shaikh, R.A.; Shashikala, S.
\newblock An Autoencoder and LSTM based Intrusion Detection Approach Against
  Denial of Service Attacks.
\newblock  In~Proceedings of~the 2019 1st International Conference on Advances in Information Technology (ICAIT), Chickmagalur, India, 25--27 July {2019}; pp.~406--410.

\bibitem[{Abolhasanzadeh}(2015)]{7288799}
{Abolhasanzadeh}, B.
\newblock Nonlinear Dimensionality Reduction for Intrusion Detection using
  Auto-Encoder Bottleneck Features.
\newblock  In~Proceedings of~the 2015 7th Conference on Information and Knowledge Technology (IKT), Urmia, Iran, 26--28 May {2015}; pp.~1--5.

\bibitem[Javaid \em{et~al.}(2016)Javaid, Niyaz, Sun, and Alam]{javaid2016deep}
Javaid, A.; Niyaz, Q.; Sun, W.; Alam, M.
\newblock A Deep Learning Approach for Network Intrusion Detection System.
\newblock  In~Proceedings of~the 9th EAI International Conference on Bio-Inspired
  Information and Communications Technologies (Formerly BIONETICS), New York, NY, USA, 3--5 December {2016}; pp.~21--26.

\bibitem[AL-Hawawreh \em{et~al.}(2018)AL-Hawawreh, Moustafa, and
  Sitnikova]{RefWorks:doc:5b0e81c9e4b01f2c3e37bf75}
AL-Hawawreh, M.; Moustafa, N.; Sitnikova, E.
\newblock Identification of Malicious Activities In Industrial Internet of
  Things Based On Deep Learning Models.
\newblock {\em J. Inf. Secur. Appl.} {\bf 2018}, {\em 41},~1--11,
\newblock
  doi:{\changeurlcolor{black}\href{https://doi.org///doi.org/10.1016/j.jisa.2018.05.002}{\detokenize{10.1016/j.jisa.2018.05.002}}}.

\bibitem[Shone \em{et~al.}(2018)Shone, Ngoc, Phai, and
  Shi]{RefWorks:doc:5ae339c4e4b0e00594a5c5cc}
Shone, N.; Ngoc, T.N.; Phai, V.D.; Shi, Q.
\newblock A Deep Learning Approach To Network Intrusion Detection.
\newblock {\em IEEE Trans. Emerg. Top. Comput. Intell.} {\bf 2018}, {\em 2},~41--50.

\bibitem[{Farahnakian} and {Heikkonen}(2018)]{8323687}
{Farahnakian}, F.; {Heikkonen}, J.
\newblock A Deep Auto-encoder Based Approach for Intrusion Detection System.
\newblock  In~Proceedings of~the 2018 20th International Conference on Advanced Communication Technology (ICACT), Chuncheon, Korea, 11--14 February {2018;} p.{~1.}

\bibitem[Meidan \em{et~al.}(2018)Meidan, Bohadana, Mathov, Mirsky, Shabtai,
  Breitenbacher, and Elovici]{meidan2018n}
Meidan, Y.; Bohadana, M.; Mathov, Y.; Mirsky, Y.; Shabtai, A.; Breitenbacher,
  D.; Elovici, Y.
\newblock N-BaIoT---Network-Based Detection of IoT Botnet Attacks Using Deep
  Autoencoders.
\newblock {\em IEEE Pervasive Comput.} {\bf 2018}, {\em 17},~12--22.

\bibitem[Bovenzi \em{et~al.}()Bovenzi, Aceto, Ciuonzo, Persico, and
  Pescap{\'e}]{bovenzihierarchical}
Bovenzi, G.; Aceto, G.; Ciuonzo, D.; Persico, V.; Pescap{\'e}, A.
\newblock A Hierarchical Hybrid Intrusion Detection Approach in IoT Scenarios.

\bibitem[{Canadian Institute for
  Cybersecurity}(2017)]{RefWorks:doc:5b227bb8e4b07f83f15ddb45}
{Canadian Institute for Cybersecurity}.
\newblock Intrusion Detection Evaluation Dataset ({CICIDS2017}). 2017.
\newblock  Available online: \url{http://www.unb.ca/cic/datasets/ids-2017.html} (accessed on: 15 June 2018). 

\bibitem[Panigrahi and Borah(2018)]{panigrahi2018detailed}
Panigrahi, R.; Borah, S.
\newblock A Detailed Analysis of {CICIDS2017} Dataset for Designing Intrusion
  Detection Systems.
\newblock {\em Int. J. Eng. Technol.} {\bf 2018}, {\em 7},~479--482.

\bibitem[Sharafaldin \em{et~al.}(2019)Sharafaldin, Habibi~Lashkari, and
  Ghorbani]{cicanalysis}
Sharafaldin, I.; Habibi~Lashkari, A.; Ghorbani, A.A.
\newblock A Detailed Analysis of the {CICIDS2017} Data Set.
\newblock In \emph{Information Systems Security and Privacy}; Mori, P., Furnell, S.,
  Camp, O., Eds.; Springer International Publishing: Cham, Switzerland, 2019; pp.~172--188.

\bibitem[{Canadian Institute for
  Cybersecurity}()]{RefWorks:doc:5b227bd4e4b0d1cffc0657b0}
{Canadian Institute for Cybersecurity}.
\newblock {NSL-KDD} Dataset.
\newblock Available online: \url{http://www.unb.ca/cic/datasets/nsl.html} (accessed on 15 June 2018).

\bibitem[Tavallaee \em{et~al.}(2009)Tavallaee, Bagheri, Lu, and
  Ghorbani]{RefWorks:doc:5d03bf33e4b0d6a9f11ef38c}
Tavallaee, M.; Bagheri, E.; Lu, W.; Ghorbani, A.A.
\newblock A Detailed Analysis of the KDD CUP 99 Data Set.
\newblock  In~Proceedings of~the 2009 IEEE Symposium on Computational Intelligence for Security and Defense Applications, Ottawa, ON, Canada, 8--10 July {2009}; pp.~1--6.

\bibitem[Tobi and Duncan(2018)]{RefWorks:doc:5bbe0843e4b0e904d9061fbb}
Tobi, A.M.A.; Duncan, I.
\newblock {KDD} 1999 Generation Faults: A Review And Analysis.
\newblock {\em J. Cyber Secur. Technol.} {\bf 2018},~1--37, doi:{\changeurlcolor{black}\href{https://doi.org/10.1080/23742917.2018.1518061}{\detokenize{10.1080/23742917.2018.1518061}}}.

\bibitem[Siddique \em{et~al.}(2019)Siddique, Akhtar, Khan, and
  Kim]{siddique2019kdd}
Siddique, K.; Akhtar, Z.; Khan, F.A.; Kim, Y.
\newblock Kdd Cup 99 Data Sets: A Perspective on the Role of Data Sets in
  Network Intrusion Detection Research.
\newblock {\em Computer} {\bf 2019}, {\em 52},~41--51.


\bibitem[Bala and Nagpal(2019)]{nslanalysis}
Bala, R.; Nagpal, R.
\newblock A Review on {KDD} {CUP99} and {NSL}-{KDD} Dataset.
\newblock {\em Int. J. Adv. Res. Comput. Sci.} {\bf 2019}, {\em 10},~64.

\bibitem[Rezaei and Liu(2019)]{rezaei2019deep}
Rezaei, S.; Liu, X.
\newblock Deep Learning for Encrypted Traffic Classification: An Overview.
\newblock {\em IEEE Commun. Mag.} {\bf 2019}, {\em 57},~76--81.

\bibitem[Guggisberg(2020)]{split}
Guggisberg, S.
\newblock How to Split a Dataframe into Train and Test Set with Python. 2020.
\newblock Available online: \url{https://towardsdatascience.com/how-to-split-a-dataframe-into-train-and-test-set-with-python-eaa1630ca7b3} (accessed on 17 August 2020).

\bibitem[Bergstra and Bengio(2012)]{bergstra2012random}
Bergstra, J.; Bengio, Y.
\newblock Random Search for Hyper-parameter Optimization.
\newblock {\em J. Mach. Learn. Res.} {\bf 2012}, {\em 13},~281--305.

\bibitem[Liashchynskyi and Liashchynskyi(2019)]{liashchynskyi2019grid}
Liashchynskyi, P.; Liashchynskyi, P.
\newblock Grid Search, Random Search, Genetic Algorithm: A Big Comparison for
  NAS.
\newblock {\em arXiv} {\bf 2019}, arXiv:1912.06059.

\bibitem[Chen \em{et~al.}(2005)Chen, Lin, and Sch{\"o}lkopf]{chen2005tutorial}
Chen, P.H.; Lin, C.J.; Sch{\"o}lkopf, B.
\newblock A Tutorial on $\nu$-Support Vector Machines.
\newblock {\em Appl. Stoch. Model. Bus. Ind.} {\bf 2005}, {\em 21},~111--136.

\bibitem[Gharib \em{et~al.}(2019)Gharib, Mohammadi, Dastgerdi, and
  Sabokrou]{gharib2019autoids}
Gharib, M.; Mohammadi, B.; Dastgerdi, S.H.; Sabokrou, M.
\newblock AutoIDS: Auto-encoder Based Method for Intrusion Detection System.
\newblock {\em arXiv} {\bf 2019}, arXiv:1911.03306.

\bibitem[Aygun and Yavuz(2017)]{aygun2017network}
Aygun, R.C.; Yavuz, A.G.
\newblock Network Anomaly Detection with Stochastically Improved Autoencoder
  Based Models.
\newblock  In~Proceedings of~the 2017 IEEE 4th International Conference on Cyber Security and Cloud Computing (CSCloud), New York, NY, USA, 26--28 June {2017; }pp.~193--198.

\end{thebibliography}
\end{document}